\documentclass[aps,preprint,nofootinbib,12pt]{revtex4}
\usepackage{amsfonts,amsthm}
\usepackage{amssymb,amsmath}
\usepackage{graphicx}

\begin{document}
\title{Gravitational energy is well defined}

\author{Chiang-Mei Chen$^{1,2}$}\email{cmchen@phy.ncu.edu.tw} \thanks{corresponding author}
\author{Jian-Liang Liu$^3$}\email{2018013@dgut.edu.cn}
\author{James M. Nester$^{1,4,5}$}\email{nester@phy.ncu.edu.tw}

\address{$^1$Department of Physics, National Central University, Chungli 32001, Taiwan}
\address{$^2$Center for High Energy and High Field Physics (CHiP), National Central University, Chungli 32001, Taiwan}
\address{$^3$Department of Mathematics and Data Science, Dongguan University of Technology, Dongguan, China}
\address{$^4$Graduate Institute of Astronomy, National Central University, Chungli 32001, Taiwan}
\address{$^5$Leung Center for Cosmology and Particle Astrophysics, National Taiwan University, Taipei 10617, Taiwan}

\date{\today}


\begin{abstract}
 The energy of  gravitating systems has been an issue since Einstein proposed general relativity: considered to be ill defined, having no proper local density.
Energy-momentum is now regarded as \emph{quasi-local} (associated with a closed 2-surface).  We consider the pseudotensor and  quasi-local proposals in the Lagrangian-Noether-Hamiltonian formulations.
There are two ambiguities: (i) many expressions, (ii) each depends on some non-dynamical structure, e.g., a reference frame.
The Hamiltonian approach gives a handle on both problems.
 Our remarkable discovery is that with a 4D isometric Minkowski reference a large class of expressions---those that agree with the Einstein pseudotensor's Freud superpotential to linear order---give a common quasi-local energy value.  With a best-matched reference on the boundary this value is the non-negative Wang-Yau mass.
\\

Essay written for the Gravity Research Foundation 2018 Awards for Essays on Gravitation.

\end{abstract}

\maketitle

A widely held opinion was expressed in an influential textbook:
\begin{quotation}\textit{Anyone who looks for a magic formula for ``local gravitational energy-momentum'' is looking for the right answer to the wrong question.  Unhappily, enormous time and effort were devoted in the past to trying to ``answer this question'' before investigators realized the futility of the enterprise.}~\cite{MTW} p 467.\end{quotation}

Here we present our recent discovery that sheds new light on gravitational energy,
 beginning with a brief review of some history.

Along with his famous field equations\footnote{Here $\kappa:=8\pi G/c^3$. Our conventions generally follow~\cite{MTW}.} $G_{\mu\nu}=\kappa T_{\mu\nu}$, whereby the density of source energy-momentum curves spacetime, Einstein also proposed~\cite{cpae} his  \emph{gravitational energy-momentum density} $\frak{t}^\mu{}_\nu$, which is \emph{non-covariantly} conserved:
\begin{equation}
\partial_\mu(\sqrt{|g|}T^\mu{}_\nu+\frak{t}^\mu{}_{\nu})=0. \label{conserved}
\end{equation}
 $\frak{t}^\mu{}_{\nu}$ is  coordinate reference frame dependent, a \emph{pseudotensor}, not a proper tensor.  Many objected, beginning with Lorentz, Levi-Civita, Felix Klein and Schr\"odinger.
 Einstein understood their concerns, but believed that his pseudotensor had physical meaning.
Emmy Noether's paper with her two famous theorems concerning symmetry in dynamical systems was written to clarify energy issues raised by Einstein, David Hilbert and Felix Klein~\cite{Kosmann-Schwarzbach}.
She conclusively showed that there is \emph{no proper conserved energy-momentum density} for any dynamical geometry gravity theory.

From (\ref{conserved}) one can infer the
existence of a \emph{superpotential} $\mathfrak{U}^{\mu\lambda}{}_\nu\equiv \mathfrak{U}^{[\mu\lambda]}{}_\nu$ satisfying
\begin{equation}
\kappa^{-1}\sqrt{|g|}G^\mu{}_\nu+\mathfrak{t}^\mu{}_\nu=\partial_\lambda \mathfrak{U}^{\mu\lambda}{}_\nu.
\end{equation}
 In 1939 Freud found such a superpotential~\cite{Freud}:\footnote{$\Gamma^\alpha_{~\beta\gamma}$ is the Levi-Civita connection.}
\begin{eqnarray}(2\kappa)\mathfrak{U}_{\rm F}^{\mu\lambda}{}_\nu&:=&-\sqrt{|g|}{g}^{\beta\sigma}
\Gamma^\alpha{}_{\beta\gamma}\delta^{\mu\lambda\gamma}_{\alpha\sigma\nu}.\label{UF}
\end{eqnarray}
Later other pseudotensors were proposed, notably by Landau-Lifshitz,
Papapetrou,
Bergmann \& Thomson,
M{\o}ller,
 and Weinberg
  (also used in~\cite{MTW}); they similarly follow from their respective superpotentials~\cite{Chang:1998wj}:
\begin{eqnarray}
2\kappa\mathfrak{U}_{\rm LL}^{\mu\lambda\nu}\!\!&:=&|\bar g|^{-\frac12}\delta^{\mu\lambda}_{\gamma\alpha}\partial_\pi[|g|
g{}^{\alpha\nu}g^{\gamma\pi}]\label{LL}
\equiv\delta^{\mu\lambda}_{\gamma\alpha}\delta^{\nu\pi}_{\beta\rho}|g/\bar g|^{\frac12}g^{\alpha\beta}
\partial_\pi(|g|^{\frac12}g^{\gamma\rho}),\\
2\kappa\mathfrak{U}_{\rm
P}^{\mu\lambda\nu}\!\!&:=&\delta^{\mu\lambda}_{\gamma\alpha}\delta^{\nu\pi}_{\beta\rho}\bar
g{}^{\alpha\beta}\partial_\pi[|g|^{\frac12}g^{\gamma\rho}]\label{HP}
\equiv\!\delta^{\mu\lambda}_{\gamma\alpha}\delta^{\nu\pi}_{\beta\rho}\bar g{}^{\alpha\beta}
|g|^{\frac12}({\textstyle\frac12}
g^{\gamma\rho} g^{\tau\delta}\!- g^{\gamma\tau} g^{\rho\delta}){\partial}_\pi g_{\tau\delta},\\
\mathfrak{U}_{\rm BT}^{\mu\lambda\nu}\!\!&:=&g^{\nu\delta}\frak{U}_{\rm
F}^{\mu\lambda}{}_\delta\equiv |\bar g/g|^{\frac12}\frak{U}_{\rm LL}^{\mu\lambda\nu},\label{UBT}\\
2\kappa\mathfrak{U}_{\rm
M}^{\mu\lambda}{}_\nu\!\!&:=&
|g|^{\frac12}\delta^{\mu\lambda}_{\alpha\sigma}g^{\beta\alpha}g^{\sigma\delta}{\partial}_\beta
g_{\delta\nu} \label{UM},\\
2\kappa\mathfrak{U}_{\rm
W}^{\mu\lambda\nu}\!\!&:=&\!\delta^{\mu\lambda}_{\gamma\alpha}\delta^{\nu\pi}_{\beta\rho}\bar g{}^{\alpha\beta}|\bar
g|^{\frac12}({\textstyle\frac12}{\bar
g}^{\gamma\rho}{\bar g}^{\tau\delta}\!-{\bar g}^{\gamma\tau}{\bar g}^{\rho\delta}){\partial}_\pi g_{\tau\delta},\ \ \ \ \label{HW}
\end{eqnarray}
where $\bar g_{\mu\nu}={\rm diag}(-1,+1,+1,+1)$, the Minkowski metric.
All define energy-momentum values which depend on the coordinate reference frame.
For meaningful values one needs Minkowski-Cartesian type coordinates.
 Two unsatisfactory issues are: (i) which expression? (ii) which reference frame?
 Here we give satisfying answers.
For now, we just note that these pseudotensors (a) do provide a description of energy-momentum conservation, (b)  do have well defined values in each reference frame, and (c) all (except  M{\o}ller~(\ref{UM})~\cite{Mol58}) give at spatial infinity the expected total energy-momentum values (described in Ref.~\cite{MTW}, \S 20.2).

One can understand the significance of these expressions
via the Hamiltonian approach.
  For any region covered by a single coordinate system choose a constant component vector field $Z^\nu$.  The total energy-momentum in the region, $P_\nu(V)$, follows from
\begin{eqnarray}
-Z^\nu P_\nu(V) &:=& - \int_V Z^\nu ({{\mathfrak{T}}^\mu{}_\nu+{\mathfrak{t}}^\mu{}_\nu}) d\Sigma_\mu
\nonumber\\
&\equiv&\!\! \int_V\! Z^\nu |g|^{\frac12} ( {\kappa}^{-1} G^\mu{}_\nu - T^\mu{}_\nu\! ) d\Sigma_\mu
\! -\! \oint_{S=\partial V} \! Z^\nu {\mathfrak{U}^{\mu\lambda}{}_\nu}{\textstyle\frac12}dS_{\mu\lambda}
 \equiv H(Z, V). \label{basicHam}
\end{eqnarray}
The superpotential determines  the {\em boundary term} 2-surface integrand.
The volume integrand is the (vanishing!) Einstein field equation---the covariant expression for the ADM Hamiltonian density (see~\cite{MTW} Ch.~21).
The value of the
pseudotensor/Hamiltonian is thus \textit{quasi-local}, determined just by the boundary term.

The modern concept is not  a local energy-momentum density, but rather  \emph{quasi-local} energy-momentum:  associated with a closed 2-surface. Pseudotensors always had this property, but this was not apparent before Freud. This property became appreciated after Penrose~\cite{Penrose} introduced the term.
According to a comprehensive quasi-local review~\cite{Sza09}, we ``have
no ultimate, generally accepted expression for the energy-momentum\dots'';
 proposed criteria for quasi-local energy-momentum expressions include: (i) vanish for Minkowski, (ii) give the standard values at spatial infinity, (iii)  non-negative energy.

The Hamiltonian variation gives information that tames the freedom in the boundary term choice: boundary conditions.  The various pseudotensor values are those of the Hamiltonian having the associated boundary conditions~\cite{Chang:1998wj}. Hence the first problem  is under control: the pseudotensor values are the values of the associated Hamiltonians, which evolve the system with their respective boundary conditions.

For Einstein's theory, from our covariant Hamiltonian formulation~\cite{Chen:1994qg} we found a set of covariant-symplectic Hamiltonian boundary terms; our preferred choice corresponds to fixing the metric  on the boundary:
\begin{equation}
2\kappa{\cal B}(Z) = \left[Z^\nu\left(\Gamma^\alpha_{~\beta\gamma}-\bar\Gamma^\alpha_{~\beta\gamma}\right)
\sqrt{|g|}g^{\beta\sigma}\delta^{\mu\lambda\gamma}_{\alpha\sigma\nu}  + \bar
D_{\beta} Z^\alpha\bigl({|g|}^{\frac12}g^{\beta\sigma}-{|\bar g|}^{\frac12}\bar g^{\beta\sigma}\bigr){\textstyle\frac12}\delta^{\mu\lambda}_{\alpha\sigma}\right]{\textstyle{\frac12}}dS_{\mu\lambda}, \label{BprefGR}
\end{equation}
where $\bar g_{\mu\nu}$ and $\bar \Gamma^\alpha{}_{\beta\gamma}$ are non-dynamic reference geometry values.  To appreciate their significance, consider measuring the geometry on the boundary of a region in order to calculate the quasi-local quantities.  If the measured values equal the reference field values then this region contains no energy-momentum; thus $\bar g_{\mu\nu},~\bar \Gamma^\alpha{}_{\beta\gamma}$ define the zero energy-momentum configuration---the vacuum state.
Here we consider only the obvious natural choice of a Minkowski geometry as  reference.

For a chosen Minkowski reference geometry on the boundary, the integral of (\ref{BprefGR}) (with $Z^\nu$ as the appropriate reference Killing vector field) defines the associated quasi-local energy-momentum (and angular momentum).
Like many other boundary term choices, for asymptotically flat spaces at spatial infinity our expressions are asymptotic to the
standard total energy-momentum and angular momentum expressions.

To fix the Minkowski reference, we proposed
(i) 4D isometric matching on the boundary 2-surface $S$, and (ii) energy optimization
as criteria for the ``best matched'' reference~\cite{Nester:2012zi}.

In a neighborhood of the 2D boundary $S$, any 4 smooth independent functions $y^i$ define a Minkowski reference:
      $ \bar g = -(dy^0)^2 + (dy^1)^2 + (dy^2)^2 + (dy^3)^2 $.
Locally this defines an embedding of a neighborhood of $S$ into  Minkowski space.
The reference metric has the components
  $ \bar g_{\mu\nu} = \bar g_{ij} y^i_{~\mu} y^j_{~\nu} $ and the reference connection (which vanishes in the $y^i$ frame) is $\bar \Gamma^\alpha_{~\beta\gamma}=x^\alpha_{~j}\partial_\gamma y^j_{~\beta}$,
  (where $dy^i=y^i_{~\mu}dx^\mu$, $dx^\mu=x^\mu_{~j}dy^j$).
With $\bar Z^\nu$ a Minkowski reference translational Killing field, the second term in~(\ref{BprefGR}) vanishes; then
   our quasi-local expression becomes
\begin{eqnarray}
2\kappa{\cal B}(\bar Z)\! &=&\!
\bar Z^k x^\nu_{~k}|g|^{\frac12} [\Gamma^\alpha{}_{\beta\gamma} - x^\alpha_{~j} \, \partial_\gamma y^j_{~\beta}] g^{\beta\sigma}\delta^{\mu\lambda\gamma}_{\nu\alpha\sigma}{\textstyle\frac12} dS_{\mu\lambda} \label{BMink}
\equiv \bar Z^k|g|^{\frac12}g^{jn}\Gamma^i_{~jl}\delta^{pql}_{kin}{\textstyle\frac12}dS_{pq}.\qquad \label{BF}\end{eqnarray}
Thus, when expressed in the Minkowski reference coordinate frame, it reduces to the Freud superpotential~(\ref{UF}).

Our first reference criterion is 4D isometric matching on the boundary $S$
(proposed by Epp in 2000~\cite{Epp:2000zr}), equivalent to embedding $S$ into Minkowski spacetime.
 To appreciate the significance, imagine making a map of an Earth surface region.    Slicing a flat 2-plane through the region gives a circular boundary along which the planar and spherical metrics exactly match.
In 4D there are  10 constraints:
    $ g_{\mu\nu}|_S = \bar g_{\mu\nu}|_S = \bar g_{ij} y^i{}_\mu y^j{}_\nu|_S $
and 12 embedding functions  on $S$ ($y^i$ and their two normal derivatives).
 Three constraints concern the 2D surface intrinsic isometric matching.
   Via a modified closed 2-surface into $\mathbb R^3$ embedding theorem, there is a unique embedding (\emph{but no explicit formula})---as long as the surface satisfies a certain convexity condition.
One can take $y^0$ and its spatial radial derivative as the embedding control variables~\cite{Nester:2012zi}.

Our surprising discovery: with a 4D isometric reference  the values of many distinct expressions coincide.
\emph{For any closed 2-surface in a dynamical Riemannian spacetime, with 4D isometric matching to a Minkowski reference, there is a
common quasi-local energy value for all the expressions that linearly agree with the Freud superpotential (\ref{UF}) in the Minkowski limit}.

It is easy to check that with an isometric Minkowski reference on $S$, all of the expressions~(\ref{UF},\ref{LL},\ref{HP},\ref{UBT},\ref{HW},\ref{BprefGR}) have the same value.
Linearly  Freud-like  expressions having a concordant value include most of the well-known pseudotensors
and many of the quasi-local expressions\footnote{All require some additional non-dynamical structure (e.g., a frame or a spinor field) that serves as the reference.}~\cite{Sza09} proposed within the Lagrangian-Noether-Hamiltonian frameworks.

In view of the quasi-local desiderata: (i) vanishing for Minkowski, (ii) agreeing with the standard spatial infinity linear results,
this concord could have been expected (even though the expressions differ beyond the linear order and disagree for other reference choices).

Some pseudotensor/Hamiltonian-boundary term expressions give other quasi-local energy values, i.e., those that do not have the desirable spatial asymptotic linearized theory limit (\cite{MTW}, \S 20.2) or do not  choose the reference by embedding into Minkowski spacetime, e.g,
M{\o}ller~\cite{Mol58}, Brown \& York~\cite{BrownYork}, Kijowski-Liu \& Yau~\cite{Kijowski97,LiuYau,Sza09}.

To fix the two embedding control variables,
one can use the boundary term value.
We  argued that the  critical points are distinguished~\cite{Nester:2012zi}, but we do not have an explicit formula for them.

From another perspective,
Wang \& Yau~\cite{WY09}
used an expression in terms of surface geometric quantities
 within the Hamilton-Jacobi approach. They analytically found the optimal reference and thereby determined their \emph{quasi-local mass};  moreover,  they were able to show that it  is \textit{non-negative} and, furthermore, \textit{vanishes for Minkowski}.
An outstanding achievement.

Recently it was found that~(\ref{BprefGR})
with a 4D isometric Minkowski reference
is closely related to the Wang-Yau expression,
with a saddle critical value of the associated energy
agreeing with the Wang-Yau mass~\cite{Liu:2017neh}.
 This is an important link, since all the linearly-like-Freud expressions with a 4D isometric Minkowski reference have the same energy value as~(\ref{BprefGR}).
Hence, (via~\cite{Liu:2017neh,WY09}) there is \textit{for all the  Freud-linear expressions} a
 non-negative quasi-local energy---which vanishes for Minkowski.

This quasi-local energy could have been found in 1939.  It was found in 2009 by Wang \& Yau~\cite{WY09}. We came to this value a few years later~\cite{Nester:2012zi}.
Einstein with his pseudotensor was close (as were many others), merely lacking a good way to choose the coordinate reference frame on the boundary.
Much effort went into finding the ``best'' expression.
Our group investigated the roles of the Hamiltonian boundary term, which led to~(\ref{BprefGR}).
Then we turned to finding a good reference.
But just taking (almost) any proposed expression and looking for the ``best'' reference could have led anyone directly to this energy.

Gravitational energy was considered to be ill defined:
(i) no unique expression,
(ii) reference frame dependent expressions with no unique reference frame.
But we find that: (a) one can generally have a 4D isometric Minkowski reference,
(b) with such a reference all of the quasi-local expressions in a large class give the same energy-momentum (and angular momentum), (c) moreover, one can find a ``best matched'' Minkowski reference; the associated energy has the desired properties.

\noindent\textbf{Acknowledgement}

C.M.C. was supported by the Ministry of Science and Technology of the R.O.C. under the grant MOST 106-2112-M-008-010. J.-L. Liu was partially supported by the National Natural Science Foundation of China 61601275.


\end{document}